\documentclass[a4paper,11pt,preprintnumbers]{article}

\pdfoutput=1

\usepackage{jheppub}
\usepackage{amsmath}
\usepackage{xspace}
\usepackage{hyperref}
\usepackage{mdwlist}

\usepackage{braket}
\allowdisplaybreaks 
\usepackage{color}
\usepackage{mathrsfs}

\newcommand{\ba}{\begin{eqnarray}}
\newcommand{\ea}{\end{eqnarray}}
\def\bea#1\eea{\begin{align}#1\end{align}}
\newcommand{\bef}{\begin{figure*}[t]\centering}
\newcommand{\eef}{\end{figure*}}
\newcommand{\be}{\begin{equation}}
\newcommand{\ee}{\end{equation}}

\def\OMIT#1{{}}

\newcommand{\GeV}{~\mathrm{GeV}}

\newcommand{\pp}{p\kern-0.05em p}
\newcommand{\PbPb}{\ensuremath{\mbox{Pb--Pb}}}

\newcommand{\Rmax}{\ensuremath{R_{\mathrm{max}}}}
\newcommand{\pT}{\ensuremath{p_{T}}}
\newcommand{\pTjet}{\ensuremath{p_{T}^{\mathrm{jet}}}}

\begin{document}

\preprint{\begin{flushright}
YITP-SB-2021-22
\end{flushright}}


\title{The information content of jet quenching and machine learning assisted observable design}

\author[a]{Yue Shi Lai,}
\author[a,b]{James Mulligan,}
\author[a]{Mateusz P\l osko\'n,}
\author[a,c,d]{Felix Ringer}

\affiliation[a]{Nuclear Science Division, Lawrence Berkeley National Laboratory, Berkeley, CA 94720, USA}
\affiliation[b]{Physics Department, University of California, Berkeley, CA 94720, USA}
\affiliation[c]{C.N. Yang Institute for Theoretical Physics, Stony Brook University, Stony Brook, NY 11794, USA}
\affiliation[d]{Department of Physics and Astronomy, Stony Brook University, Stony Brook, NY 11794, USA}

\emailAdd{ylai@lbl.gov}
\emailAdd{james.mulligan@berkeley.edu}
\emailAdd{mploskon@lbl.gov}
\emailAdd{felix.ringer@stonybrook.edu}


\abstract{ 
Jets produced in high-energy heavy-ion collisions are modified compared to those in proton-proton collisions due to their interaction with the deconfined, strongly-coupled quark-gluon plasma (QGP).
In this work, we employ machine learning techniques to identify important features that distinguish jets produced in heavy-ion collisions from jets produced in proton-proton collisions.
We formulate the problem using binary classification and focus on leveraging machine learning in ways that inform theoretical calculations of jet modification: (i) we quantify the information content in terms of Infrared Collinear (IRC)-safety and in terms of hard vs. soft emissions,
(ii) we identify optimally discriminating observables that are in principle calculable in perturbative QCD, and 
(iii) we assess the information loss due to the heavy-ion underlying event and background subtraction algorithms.
We illustrate our methodology using Monte Carlo event generators, where we find that important information about jet quenching is contained not only in hard splittings but also in soft emissions and IRC-unsafe physics inside the jet.
This information appears to be significantly reduced by the presence of the underlying event. We discuss the implications of this for the prospect of using jet quenching to extract properties of the QGP. 
Since the training labels are exactly known, this methodology can be used directly on experimental data without reliance on modeling.
We outline a proposal for how such an experimental analysis can be carried out,
and how it can guide future measurements.}

\maketitle

\newpage

\section{Introduction \label{sec:intro}}

Jets are highly energetic and collimated groups of particles observed in the detectors of high-energy scattering experiments such as the Large Hadron Collider (LHC) and the Relativistic Heavy Ion Collider (RHIC). Jets reflect the underlying quark and gluon degrees of freedom of hard-scattering processes, in which the scattered partons fragment into a shower due to multiple soft and collinear emissions. 
These partons subsequently hadronize into the color neutral particles observed in detectors. Studies of the radiation pattern inside jets, known as jet substructure, aim to quantify and utilize jets to better understand quantum chromodynamics (QCD) and search for physics beyond the Standard Model~\cite{Larkoski:2017jix,Asquith:2018igt,Marzani:2019hun}. Jets and their substructure have been studied both in proton-proton ($pp$) and heavy-ion ($AA$) collisions. 

In high-energy heavy-ion collisions a state of matter known as the quark-gluon plasma (QGP) is formed, in which quarks and gluons are deconfined~\cite{Busza:2018rrf}. 
The QGP is believed to have populated the universe for most of the first few microseconds after the Big Bang \cite{Weinberg:1977ji}, and today serves as a laboratory to study the emergence
of a high temperature, strongly-coupled system from quarks and gluons in QCD.
While several of its transport coefficients have been constrained by experimental measurements, the detailed properties of the QGP remain unknown. Information about the QGP can be obtained by comparing jets in vacuum to their counterparts in heavy-ion collisions which have traversed the hot and dense nuclear matter. The modifications of the jets emerging after interacting with the QGP as compared to jets in $pp$ collisions is known as \emph{jet quenching}. In an ongoing effort to achieve a unified description of jet modification a significant experimental and theoretical effort has been made to measure and compute the medium-induced modifications of various jet observables  ~\cite{Gyulassy:1993hr,Baier:1996sk,Zakharov:1996fv,Gyulassy:2000er,Wang:2001ifa,Arnold:2002ja,Liu:2006ug,Ovanesyan:2011xy,Armesto:2011ht,Mehtar-Tani:2013pia,Burke:2013yra,Brewer:2018dfs,Qiu:2019sfj,JETSCAPE:2021ehl,Caucal:2019uvr,Andres:2020vxs,Vaidya:2020cyi,Vaidya:2021mly,Clayton:2021uuv, ATLAS:2018gwx,ALICE:2019qyj,CMS:2021vui, ALICE:2021obz}. 

Several guiding principles have been used to identify suitable observables in order to extract properties of the QGP. First, observables insensitive to collinear and soft emissions are given prominence -- the so-called Infrared Collinear (IRC) safe observables. IRC safety (or by extension Sudakov safety~\cite{Larkoski:2013paa,Larkoski:2015lea,Procura:2018zpn}) ensures that they can be calculated in perturbation theory order-by-order and that they are less sensitive to experimental resolution effects.
While nonperturbative effects can still be important, IRC safe observables do not require nonperturbative quantities such as parton-to-hadron fragmentation functions. Often observables that are under good theoretical control and those with a relatively simple factorization structure are preferred. This allows for precision computations to high perturbative accuracy and in several cases it allows for predictions on how nonperturbative contributions scale~\cite{Dokshitzer:1995zt,Korchemsky:1999kt,Lee:2006nr,Dasgupta:2007wa,Makris:2017arq,Hoang:2019ceu,Cal:2019gxa,Chen:2020vvp}. A second group of guiding principles for selecting an observable stems from experimental considerations where observables that can be measured well in the large fluctuating background in heavy-ion collisions are favored~\cite{ALICE:2015mdb, CMS:2021vui, Mulligan:2020tim}. Third, observables are chosen based on their sensitivity to specific medium effects such as energy loss or $p_T$-broadening. 
This can be studied in the context of Bayesian inference (see e.g. Ref~\cite{JETSCAPE:2020mzn}),
or by using machine learning to automate the design of observables that are most sensitive to model parameters such as the temperature profile of hydrodynamic simulations~\cite{Lai:2018ixk}. In this work, we explore a new guiding principle: the systematic quantification of the relative information content of quenched vs. vacuum jets.

In recent years machine learning techniques have been applied to studies in nuclear and high-energy physics~\cite{deOliveira:2015xxd,Komiske:2016rsd,Pang:2016vdc,Monk:2018zsb,Larkoski:2019nwj,Kasieczka:2020nyd,Lai:2020byl,Bieringer:2020tnw,Bernreuther:2020vhm,Huang:2021iux,Andreassen:2018apy,Alanazi:2020jod,Faucett:2020vbu,Dreyer:2020brq,H1:2021wkz,Ghosh:2021roe,Konar:2021zdg, Dillon:2021gag}.
In particular, several publications have applied machine learning methods to study jet quenching~\cite{Chien:2018dfn, Lai:2018ixk, Du:2021pqa, Apolinario:2021olp}.
Yet, one of the outstanding challenges in the application of machine learning to jet quenching is model dependence: supervised machine learning algorithms require training data, and when relying on Monte Carlo event generators to produce the training input, it is often unclear how to connect the results to jet quenching observed in nature.

\begin{figure}[t]
\centering
\includegraphics[scale=0.4]{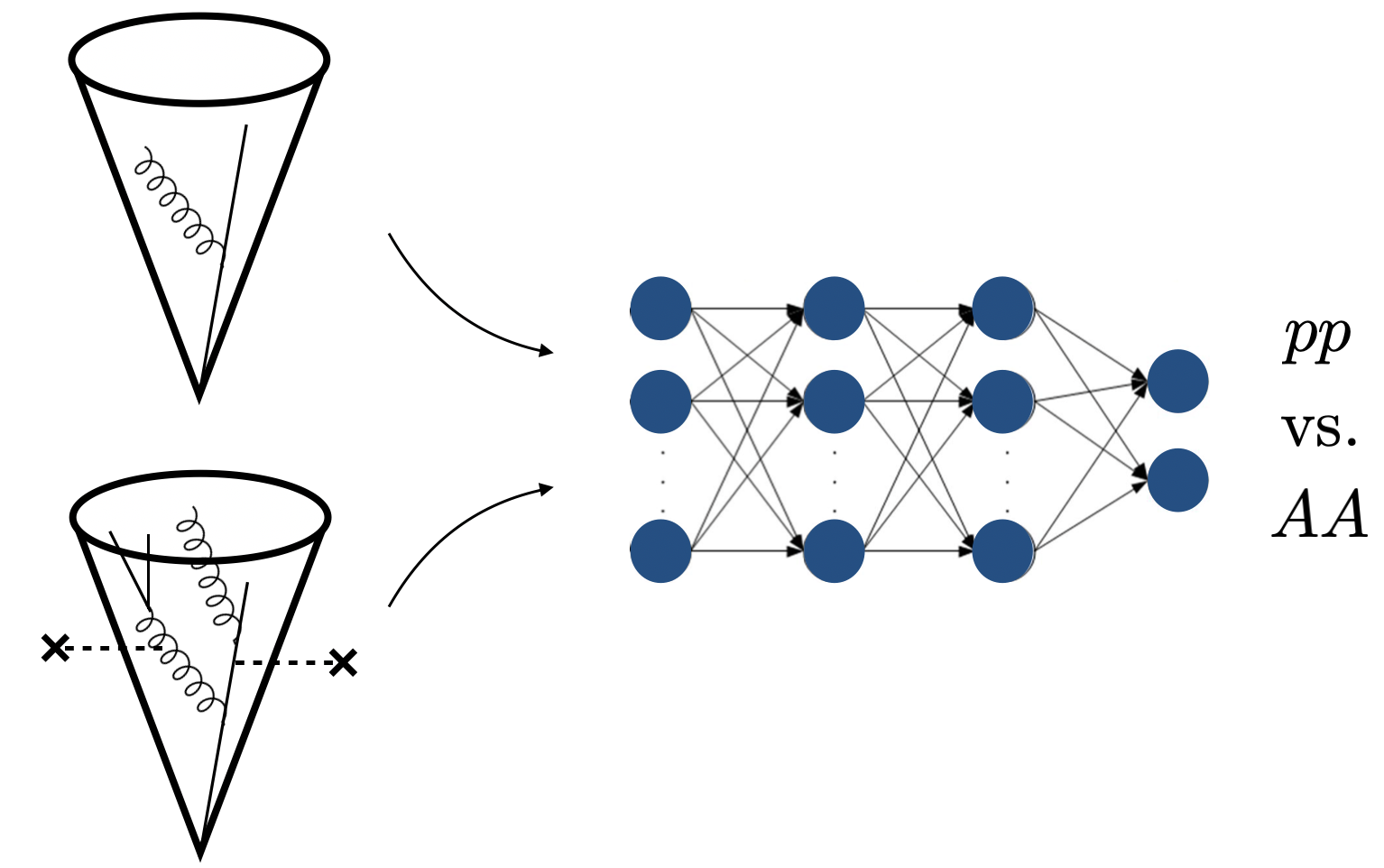}
\caption{Schematic illustration of jets in $pp$ (left) and heavy-ion $AA$ (right) collisions. Interactions with the QGP can lead to a modification of the jet substructure. By training a binary classifier, the machine learns the relevant information that distinguishes jets in $pp$ and $AA$ collisions.~\label{fig:pp_AA}}
\end{figure}

In this work, we examine jet quenching as a classification problem – distinguishing jets in proton-proton collisions from jets in heavy-ion collisions, as illustrated in Fig.~\ref{fig:pp_AA}.
We utilize machine learning to quantify the features and patterns that distinguish these two classes of jets.
In the following we refer to the set of identified features as the \emph{information content of jet quenching}.
Rather than training the classifier purely as a black box, we focus on formulating the classifier in a way that allows us to dissect the machine-learned information in order to provide insights about jet quenching. In particular, we curate the input variables and the theoretical properties of the machine learning architecture in order to not only to estimate an upper bound on the information content in jet quenching but also to study how that information is distributed in terms of IRC safety and hard vs. soft physics.
We follow this up by performing Lasso regression~\cite{doi:10.1137/0907087,10.2307/2346178} (a simple form of symbolic regression) to design jet substructure observables that are maximally modified by the QGP. 
Strongly modified observables can provide guidance of how theoretical formulations of jet quenching should modify jet evolution.
We identify single observables that are in principle calculable in perturbative QCD, by which we mean observables with closed-form expressions that obey IRC-safety or Sudakov safety and therefore can generally be calculated in the future with existing methods.
This approach provides a handle between discrimination power and simplicity of the observable which allows for a bridge between machine-learned information and quantities that have the potential to be understood in perturbative QCD.
With this method, we demonstrate that the discrimination power of our designed observables exceeds that of traditional observables already measured in heavy-ion collisions. 
We also quantify the information loss due to the large underlying event, one of the most significant challenges in jet quenching.
For the first time, 
we study the information loss caused by background subtraction techniques that are traditionally employed in heavy-ion collisions.

To illustrate this methodology, we perform binary classification tasks using Monte Carlo event generators. We expect that the proposed analyses can and should ideally be carried out using event-by-event data from the experiments at the LHC and RHIC.
These analyses can be performed either using uncorrected experimental data or using corrected full events~\cite{Andreassen:2019cjw}; these experimental considerations will be discussed further in Section~\ref{sec:experiment}. 
Unlike other classification tasks in high-energy and nuclear physics, such as quark vs. gluon jet tagging, 
our goal is to classify two experimentally measurable distributions rather than to directly relate the measured jet properties to parton level quantities,
and so there are no ambiguities concerning the labels of the classification task. 
One is therefore able to use fully supervised learning without reliance on modeling.
We expect that the proposed data driven approach will serve as a guiding principle to construct observables in heavy-ion collisions and complement existing approaches in the literature.
In addition, our approach can be useful for the extraction of medium parameters by extending the existing set of observables to new ones which contain valuable additional information~\cite{Cao:2021keo}.
We also note that while we perform our analysis using only the particles inside identified inclusive jets, similar results can be obtained using full event information or other jet topologies. It can also be applied to (jets or events) in $ep$ vs. $eA$ collisions at the future Electron-Ion Collider (EIC)~\cite{AbdulKhalek:2021gbh}.

The remainder of our paper is organized as follows. In Section~\ref{sec:event_generation}, we discuss the event generation of the jet samples in $pp$ and $AA$ collisions and the treatment of the heavy-ion underlying event. In Section~\ref{sec:classification},
we present results examining IRC-safe information (Section~\ref{sec:ML_PFN}) and soft physics via a jet substructure basis (Section~\ref{sec:ML_IRC_set}) using different machine learning architectures. 
In Section~\ref{sec:lasso}, we use a form of symbolic regression to design observables that approximate these classifiers.
In Section~\ref{sec:background}, we study the impact of the heavy-ion underlying event on jet classification. In Section \ref{sec:experiment}, we discuss experimental considerations which are relevant in order to perform the proposed analysis on actual data instead of parton shower simulations. We conclude and present an outlook in Section~\ref{sec:conclusions}.

\section{Monte Carlo jet sample ~\label{sec:event_generation}}

To illustrate our methodology, we perform our analysis using jets from Monte Carlo event generators at $\sqrt{s_{\rm{NN}}}=5.02$ TeV. To generate vacuum jets, we use PYTHIA8~\cite{Sjostrand:2007gs} with the Monash2013 tune~\cite{Skands:2014pea}. To generate medium modified jets, we use JEWEL~\cite{Zapp:2012ak,Zapp:2013vla} 2.2.0 with an initial temperature $T_i=590$ MeV and initial quenching time $\tau_i=0.4$ (formation time of the QGP), which provides an adequate description of a variety of jet quenching observables~\cite{KunnawalkamElayavalli:2017hxo}.
We do not include recoil particles in JEWEL, both for simplicity and
because the physics of medium response is relatively poorly understood; we defer this 
to a future study using experimental data.
In PYTHIA8, we disable multi-parton interactions (MPI) in order to match JEWEL.

In both cases, we reconstruct inclusive jets with the anti-k$_T$ algorithm~\cite{Cacciari:2008gp} and $R=0.4$. We consider jets with transverse momentum \pTjet{} in the range $100<\pTjet<125\;\GeV$ and pseudorapidity of $|\eta^{\mathrm{jet}}| < 2$.
We neglect particle identification (PID) and assign a pion mass assumption to all particles, leaving a study of the information conveyed by PID to future work.
The \pTjet{} range is determined from the final state particles, in both PYTHIA8 and JEWEL. 

We choose to consider an inclusive jet sample because inclusive jets are the most straightforward to measure and theoretically well understood -- the initial hard-scattering distribution is fixed by the quantities ($\pTjet,\eta^{\mathrm{jet}},\sqrt{s_{\rm NN}}$) and, hence, the allowed range of the initial parton transverse momentum $\hat p_T^{\rm jet}$~\cite{Dasgupta:2014yra,Kaufmann:2015hma,Kang:2016mcy,Dai:2016hzf}. Our studies could be extended to disentangle medium quark/gluon fractions and jet-by-jet modifications using $\gamma/Z$-tagged jets and di-jets (which require a more complicated factorization to connect to the initial hard scattering distribution), which add sensitivity to the surface bias in heavy-ion collisions~\cite{Renk:2012ve}), along with additional quark/gluon taggers or theoretical constraints.

In Sections \ref{sec:classification} and \ref{sec:lasso}, we consider only the events generated by PYTHIA8 and JEWEL, which contain no underlying event stemming from particle production in uncorrelated scatterings in heavy-ion collisions.
In Section \ref{sec:background}, we study the impact of the heavy-ion underlying event by 
combining the particles from an event simulated by PYTHIA8 or JEWEL together with the particles from a randomly drawn simulated background event.
We generate a heavy-ion background
based on a thermal model consisting of $N$ particles drawn from a Gaussian with 
$\left< {\rm d}N/{\rm d}\eta\right> \approx 2500 $ and \pT{} sampled event-by-event from a Gamma distribution, 
$f_\Gamma \left( \pT;\alpha,\beta \right) \propto \pT^{\alpha-1} e^{-\pT/\beta}$
with $\alpha=2$. We select $\beta=0.4$, corresponding to $\left< \pT \right>\approx 0.8$~GeV, in order to approximate the particle multiplicities, fluctuations, and transverse momentum 
observed in 0-10\% central \PbPb{} data \cite{Abelev2012, Mulligan:2020tim}.

After combining the PYTHIA8 or JEWEL particles together with the background particles, 
we make use of a background subtraction procedure to mitigate the influence of the large combinatorial background on reconstructed observables, as done in experimental analyses.
We employ event-by-event constituent subtraction, which corrects the
overall jet \pT{} and its substructure simultaneously by subtracting
the underlying event energy constituent-by-constituent~\cite{Berta:2014eza, Berta:2019hnj}.
In each event the subtraction algorithm distributes a uniform density of soft particles according to the average background density in the event, and subtracts their \pT{} from the original particles in the event within a 
 maximum recombination rapidity-azimuth distance \Rmax{}.
We use a value of $\Rmax=0.25$, with other parameters set to their default values.
We then perform jet reconstruction in the same manner as above, except that the 
\pTjet{} range is selected based on the particles from PYTHIA8 or JEWEL alone, in order to allow comparison to studies without background.

We use a sample of approximately $2\times 10^5$ jets, balanced evenly between $pp$ and $AA$.
These jet samples will serve as input to several different machine learning architectures described in the following sections.
We reserve 20\% of the training sample as a validation set, and an additional 20\% as a test set.
The size of the sample is determined by studying when the classification performance approximately saturates as the statistics are increased (noting that our model-dependent study is not meant to achieve sub-percent level optimization), and corresponds roughly to the number of jets recorded with an integrated luminosity of $\mathcal{O}(1\;\rm{fb}^{-1})$ at the LHC energies.
In all cases, we report classification metrics on the test set, and ensure compatible performance with the training and validation sets.
In general, it is sufficient to train a per-instance (jet-by-jet) instead of a per-ensemble classifier since collider events are independently and identically distributed~\cite{Nachman:2021yvi}.

The performance of a classifier is typically assessed by analyzing the receiver operating characteristic (ROC) curve and the area under the ROC curve (AUC). The ROC curve shows the cumulative distribution functions of the true positive rate vs. the false positive rate as the decision threshold is varied. In our case we define ``positive'' to refer to $AA$ jets. A random classifier follows a diagonal line with AUC$=0.5$ and the better a classifier is, the closer the curve is to the upper left edge of the plot, with a perfect classifier having AUC$=1$.

\section{Jet classification: pp vs. AA \label{sec:classification}}

The information content of jet quenching can be studied as a binary classification problem of jets in $pp$ vs. $AA$ collisions.
This classification task can be tackled with machine learning techniques: by evaluating the performance of the trained classifiers, we are able to quantify how much information the machine has learned to distinguish the two samples of $pp$ and $AA$ jets. We will explore two sets of machine learning architectures: the first to quantify the IRC-safe vs. IRC-unsafe information, and the second to quantify the information content in hard vs. soft physics. The latter approach will be the starting point for the machine-learning assisted observable design discussed in Section~\ref{sec:lasso} below.
In this Section, we consider only the jets generated by PYTHIA8 and JEWEL, without any background present from the heavy-ion underlying event.
While the results in this section are presented using Monte Carlo event generators, they can be performed on either uncorrected experimental data or corrected full events; we outline a proposal for how such an experimental analysis can be carried out in Section~\ref{sec:experiment}.

First, we will use a classifier based on deep sets~\cite{DBLP:journals/corr/ZaheerKRPSS17,DBLP:journals/corr/abs-1901-09006,JMLR:v21:19-322} or Particle Flow Networks (PFNs)~\cite{Komiske:2018cqr} which take as input sets of jet-by-jet four vectors. Since PFNs include the four vectors of all soft and collinear particles, they can make use of IRC unsafe information. Second, we compare the PFN-based classifier to results based on Energy Flow Networks (EFNs) which are an IRC-safe analogue of PFNs~\cite{Komiske:2018cqr}. The difference in performance of these two classifiers can be considered as a measure of the IRC unsafe information content of quenched jets. 

To examine the information content in hard vs. soft physics, we use a complete set of IRC safe observables which spans the phase space of a jet's substructure. We explore two complete sets. The first is based on the approach developed in Refs.~\cite{Datta:2017rhs,Datta:2017lxt,Datta:2019ndh} using the $N$-subjettiness observables~\cite{Thaler:2010tr,Thaler:2011gf}, which we feed as input to a Dense Neural Network (DNN). See also Refs.~\cite{Stewart:2010tn,Larkoski:2015uaa,Napoletano:2018ohv,Dasgupta:2021kgi}. The second complete set of observables we explore is based on Energy Flow Polynomials (EFPs)~\cite{Komiske:2017aww}. 
EFPs constitute a linear basis of jet substructure observables and, correspondingly, we train a linear classifier. 
We then study the discrimination power of the classifier as a function of the number of basis elements, and examine when the discrimination power saturates, indicating that all the relevant information is captured.

\subsection{IRC-safe vs. IRC-unsafe information content~\label{sec:ML_PFN}}

We start by reviewing permutation invariant neural network architectures. We consider both the IRC-unsafe PFNs and the IRC-safe EFNs~\cite{Komiske:2018cqr}. We then train PFN/EFN-based classifiers on the $pp$ vs. $AA$ jet samples introduced above.

\subsubsection{Particle and Energy Flow Networks~\label{sec:PFN_EFN}}

Deep sets~\cite{DBLP:journals/corr/ZaheerKRPSS17,DBLP:journals/corr/abs-1901-09006,JMLR:v21:19-322} allow for a permutation invariant representation of data and naturally accommodate data sets with variable length. They are well suited for collider physics applications where deep sets can be trained directly on event-by-event particle four vectors. In the context of collider physics, deep sets were introduced in Ref.~\cite{Komiske:2018cqr} and referred to as PFNs, and subsequently studied with related architectures~\cite{Qu:2019gqs,Dolan:2020qkr,ATLAS:2020jip,DBLP:journals/corr/abs-1910-02421,Romero:2021qlf,Mikuni:2021pou,Ostdiek:2021bem}. 

Consider a classifier represented by the function $f$, which depends on the four momenta $p_i$ ($i=1,\ldots,M$) of $M$ particles in the jet. In order to avoid that the classifier learns an unphysical ordering, it needs to be permutation invariant with respect to the per-event input variables, i.e. $f(p_1,\ldots,p_M)=f(p_{\pi(1)},\ldots,p_{\pi(M)})$, where $\pi$ is the permutation operator. Following Ref.~\cite{DBLP:journals/corr/ZaheerKRPSS17}, we can approximate the function $f$ as
\begin{equation}\label{eq:PFN}
    f\left(p_{1}, \ldots, p_{M}\right)=F\left(\sum_{i=1}^{M} \Phi\left(p_{i}\right)\right) \,,
\end{equation}
where the functions $\Phi,F$ represent neural networks. The function $\Phi$ maps the momenta $p_i$ to a $d$-dimensional latent space, where $d$ can be chosen to be sufficiently large such that it approximates the function $f$ arbitrarily well. We then sum over the functions $\Phi(p_i)$, in order to ensure that the result is permutation invariant. Finally, the function $F$ is a map from the $d$-dimensional latent space to the scalar value of the function $f$ from which the loss is calculated. 

To separate the IRC safe information from the IRC unsafe information, we consider the EFNs of Ref.~\cite{Komiske:2018cqr}.
IRC safety can be built into the permutation invariant neural network in Eq.~(\ref{eq:PFN}) as follows. Every particle inside a jet can be written in terms of its transverse momentum $p_{Ti}$ (relative to the beam axis), rapidity $y_i$ and azimuthal angle $\phi_i$. We normalize the transverse momenta to the total jet transverse momentum and obtain the dimensionless momentum fractions $z_i=p_{Ti}/\sum_j p_{Tj}$. In addition, we introduce a 2-component vector which contains the angular variables $\hat p_i=(y_i,\phi_i)$. The corresponding permutation invariant and IRC safe neural networks, $\tilde f$, can then be written as
\begin{equation}\label{eq:EFN}
    \tilde f\left(p_{1}, \ldots, p_{M}\right)=F\left(\sum_{i=1}^{M} z_i\Phi\left(\hat p_{i}\right)\right) \,.
\end{equation}
Due to the weighting of $\Phi$ with the momentum fraction $z_i$, the resulting expression is IRC safe~\cite{Komiske:2018cqr}.

We parametrize the functions $\Phi$ and $F$ in Eqs.~(\ref{eq:PFN}) and~(\ref{eq:EFN}) in terms of DNNs, using the \texttt{EnergyFlow} package~\cite{Komiske:2018cqr} with \texttt{Keras}~\cite{chollet2015keras}/\texttt{TensorFlow}~\cite{tensorflow2015-whitepaper}. For $\Phi$ we use two hidden layers with 100 nodes each and a latent space dimension of $d=256$. For $F$ we include three layers with 100 nodes each. For each dense layer we use the ReLU activation function~\cite{nair2010rectified} and we use the softmax activation function for the final output layer of the classifier. We train the neural networks using the Adam optimizer~\cite{Kingma2015AdamAM} with learning rates ranging from $10^{-3}$ to $10^{-4}$. We use the binary cross entropy loss function~\cite{https://doi.org/10.1111/j.2517-6161.1958.tb00292.x}, and train for 10 epochs with a batch size of 500. We find no significant changes in performance when changing the size of the layers, latent space dimension, learning rate, and batch size by factors of 2-10.

For each reconstructed jet, we record the transverse momentum, rapidity and azimuthal angle $(p_{Ti},y_i,\phi_i)$ of each particle $i$ inside the jet. Following Ref.~\cite{Komiske:2018cqr}, we perform a preprocessing step to simplify the training process. We rescale the transverse momenta of each particle inside the jet with the total transverse momentum of the observed jet. In addition, we center the rapidity and azimuthal angles of the particles in the jet with respect to the jet direction. The jet axis is determined using the $E$-scheme~\cite{Blazey:2000qt}. Here we only consider PFNs without PID and we leave a more detailed exploration for future work. We benchmark our setup 
using the quark- vs. gluon-jet data set provided in Ref.~\cite{Zenodo:EnergyFlow:Pythia8QGs} as well as our own generated quark and gluon samples with PYTHIA8, finding compatible results with Ref.~\cite{Komiske:2018cqr}.

\begin{figure*}[!t]
\centerline{
\includegraphics[width = 0.8\textwidth]{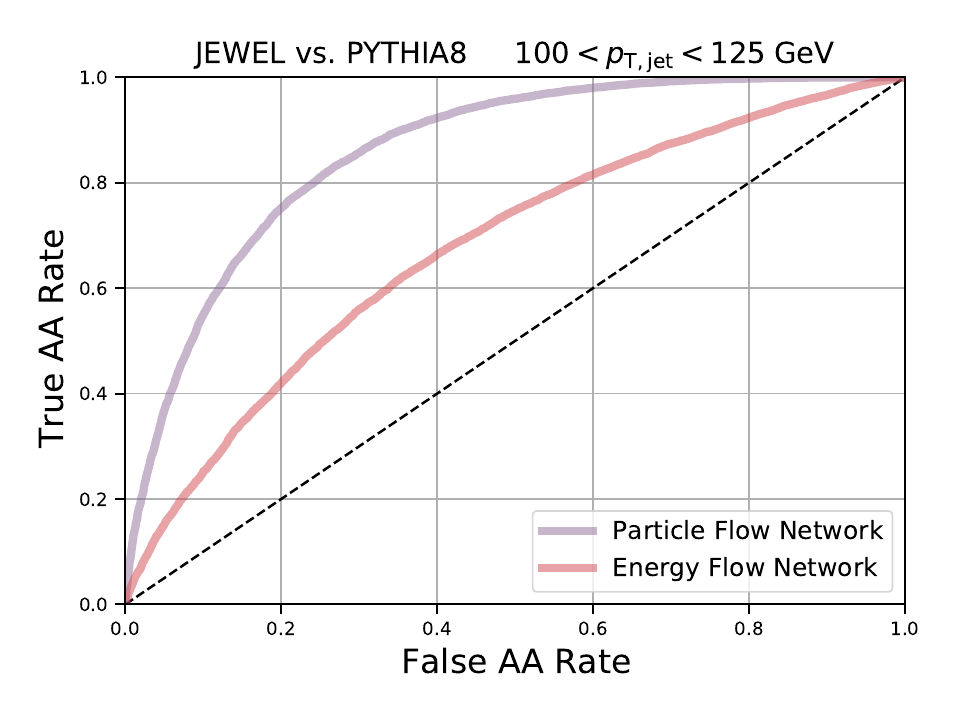}}
\caption{Classification performance of $pp$ vs. $AA$ jets quantified in terms of ROC curves using IRC-unsafe PFNs and IRC-safe EFNs. The jet samples in $pp$ and $AA$ collisions are obained from Pythia~8~\cite{Sjostrand:2007gs} and Jewel~\cite{Zapp:2012ak,Zapp:2013vla}.~\label{fig:deepset_ROC}}
\end{figure*}

Figure~\ref{fig:deepset_ROC} shows the ROC curve for $pp$ vs. $AA$ jets using the PFNs and EFNs. The AUC is $0.860$ for the PFN and $0.675$ for the EFN. Since PFNs can efficiently make use of all the available information, we use them as a benchmark for the other classification techniques discussed below. 

The difference in the classification performance between PFNs and EFNs can be considered as a measure of the IRC unsafe information contained in jet quenching.
We observe that the classifier based on IRC-unsafe PFNs significantly outperforms the one based on IRC-safe EFNs. While our findings here are model dependent, this result indicates that a significant part of the information content of jet quenching is in the IRC unsafe physics. This striking difference between PFNs and EFNs was not observed for quark/gluon jet tagging in Ref.~\cite{Komiske:2018cqr}. 
Our findings suggest that in order to advance our understanding of the quenching process, it will be valuable to measure jet substructure observables in heavy-ion collisions that go beyond IRC-safety. If model parameters such as $\hat q$ were extracted only using IRC-safe quantities such as the jet mass, a large amount of additional information which is encoded in the modified jet substructure in heavy-ion collisions would be missed. 
Examples of IRC-unsafe observables that are calculable within 
perturbative QCD include Sudakov safe observables~\cite{Larkoski:2013paa,Larkoski:2015lea,Procura:2018zpn},
such as the groomed momentum sharing fraction~\cite{Larkoski:2015lea,Cal:2021fla}, ratios of jet angularities~\cite{Larkoski:2013paa,Procura:2018zpn}, and the jet pull~\cite{Gallicchio:2010sw,Larkoski:2019urm};
or hadronic observables, which require the introduction of nonperturbative components, such as the transverse momentum spectra of hadrons~\cite{Collins:1989gx} or the jet charge~\cite{Krohn:2012fg,Waalewijn:2012sv}.
Our results can therefore provide guidance for future efforts to explore which IRC-unsafe observables complement existing results and allow us to extract the maximum information content of jet quenching.

\subsection{Hard vs. soft information content~\label{sec:ML_IRC_set}}

In this section, we consider a different data representation and machine learning setup based on complete sets of IRC safe observables. 
In order to examine the distribution of information, we train classifiers using IRC safe observables as input and we explore when the discrimination power of the classifier saturates as more observables are included.

The number of jet substructure elements required to saturate the information content depends in general on the jet substructure basis employed. 
We consider two different choices.
In Section~\ref{sec:Nsub}, we employ the $N$-subjettiness basis introduced in Ref.~\cite{Datta:2017rhs}. 
The $N$-subjettiness basis offers the benefit of minimally describing the available phase space, and vanishes when $N$ exceeds the particle multiplicity. 
In Section~\ref{sec:EFP}, we consider EFPs which were introduced in Ref.~\cite{Komiske:2017aww} as a linear IRC safe basis of jet substructure. Through their linear nature, EFPs are well suited for machine learning assisted observable design discussed in Section~\ref{sec:lasso}. Individual EFPs and certain linear combinations are calculable in perturbative QCD. See Ref.~\cite{Cal:2022fnm} for recent progress in this direction.

\subsubsection{$N$-subjettiness basis~\label{sec:Nsub}}

\begin{figure}[t]
\centering
\includegraphics[width = 0.9 \textwidth]{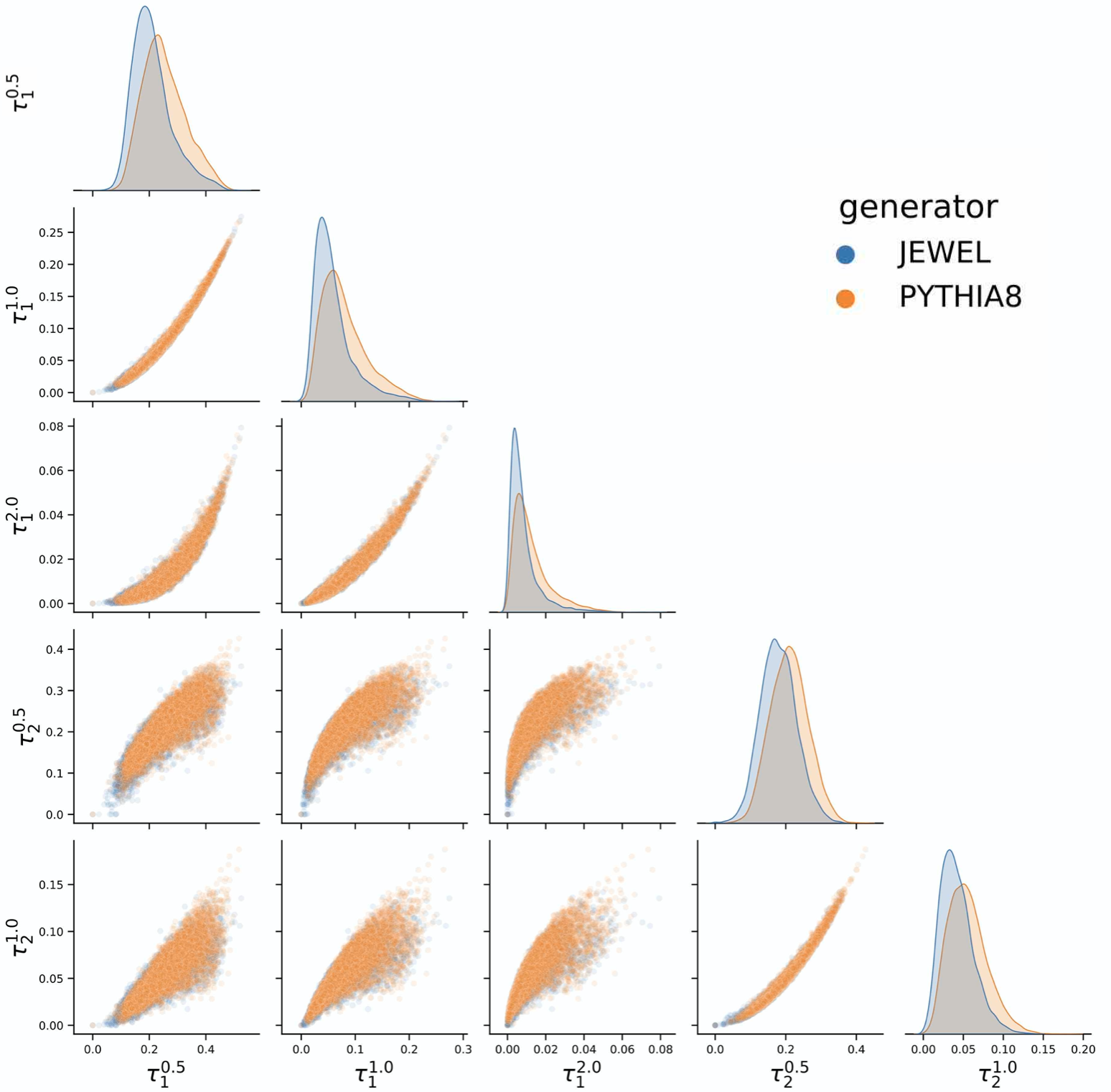}
\caption{Scatter plot showing different $N$-subjettiness distributions (diagonal) and their pairwise correlations (off-diagonal panels) in $pp$ and $AA$ collisions without background. The $pp$ and $AA$ results shown here are obtained from Pythia~8~\cite{Sjostrand:2007gs} and Jewel~\cite{Zapp:2012ak,Zapp:2013vla}, respectively.~\label{fig:nsubjettiness}}
\end{figure}

We start by reviewing the $N$-subjettiness observables~\cite{Thaler:2010tr,Thaler:2011gf,Stewart:2010tn,Napoletano:2018ohv} which provide a complete and minimal basis of the $M$-body phase space of emissions inside the jet~\cite{Datta:2017rhs}. The $N$-subjettiness observables $\tau_N^{(\beta)}$ are defined as follows. First, we identify $N$ axes inside the jet using the exclusive $k_T$ algorithm~\cite{Ellis:1993tq,Catani:1993hr} with the $E$-recombination scheme~\cite{Blazey:2000qt}. The $N$-subjettiness variables measure the radiation in the direction of these axes:
\begin{equation}
    \tau_{N}^{(\beta)}=\frac{1}{\pTjet} \sum_{i \in \mathrm{Jet}} p_{T i} \min \left\{R_{1 i}^{\beta}, R_{2 i}^{\beta}, \ldots, R_{N i}^{\beta}\right\} \,,
\end{equation}
where $p_{Ti}$ is the transverse momentum of particle $i$ inside the jet and $R_{ji}$ is its distance in the $\eta-\phi$ plane with respect to the identified axes $j$. The exponent $\beta>0$ is a tuneable parameter. For a given $N$ and $\beta$, the $N$-subjettiness observables output a single number per jet which quantifies the radiation pattern inside the jet. We note that for $\leq N$ particles inside the jet, the $N$-subjettiness observables vanish. We compute the $N$-subjettiness observables using the implementation in the FastJet Contrib software package~\cite{Cacciari:2011ma}. Figure~\ref{fig:nsubjettiness} shows the $N$-subjettiness distributions for several different $N$ and $\beta$ values for jets in both $pp$ and $AA$ collisions (diagonal panels) and their pairwise correlation (off-diagonal panels).

Following Ref.~\cite{Datta:2017rhs}, we consider $3M-4$ $N$-subjettiness observables per jet which completely specify the $M$-body phase space. For example, the 2-body phase space is $3\cdot 2-4=2$ dimensional and the measurement of two $1$-subjettiness variables with different $\beta$ values, e.g. $\tau_1^{(1)}$ and $\tau_1^{(2)}$, is sufficient to completely specify the kinematics of the two emissions inside the jet, equivalent to the longitudinal momentum fractions $z,1-z$ and the relative angle $\theta$ of the two emissions. 
For each additional emission, three phase space dimensions are added which are fully specified by measuring three additional $N$-subjettiness observables. We use the following observables to specify the $M$-body phase space~\cite{Datta:2017rhs}:
\begin{align}
    &\text{2-body}:\;\tau_1^{(1)},\tau_1^{(2)} \,,\nonumber\\
    &\text{3-body}:\;\tau_1^{(0.5)},\tau_1^{(1)},\tau_1^{(2)},\tau_2^{(1)},\tau_2^{(2)} \,,\nonumber\\
    & \quad \; \vdots \nonumber\\
    &\text{$M$-body}:\; \tau_1^{(0.5)},\tau_1^{(1)},\tau_1^{(2)},\ldots,\tau_{M-2}^{(0.5)},\tau_{M-2}^{(1)},\tau_{M-2}^{(2)},\tau_{M-1}^{(1)},\tau_{M-1}^{(2)} \,.
\end{align}

We note that the $M$-body phase space discussed here does not directly correspond to individual particles inside the jet but rather to ``resolved emissions'' inside the jet. This is analogous to the Soft Drop procedure~\cite{Larkoski:2014wba} where a sufficiently hard emission passes the Soft Drop criterion. 
In an IRC safe representation of a jet, soft and collinear particles are clustered first leaving only the $M$ most resolved emissions, which correspond to the $3M-4$ $N$-subjettiness observables passed to the machine learning algorithm.
If $M$ is sufficiently large, the kinematics of every particle can be fully reconstructed from the $N$-subjettiness observables. 
While the $N$-subjettiness observables are all IRC safe, nonperturbative effects can nevertheless play an important role. Especially the observables with large values of $N$ are sensitive to very low energy scales. By varying the value of $M$ we can control the sensitivity to soft and collinear emissions inside the jet.\footnote{We thank Andrew Larkoski for clarifying discussions.}

We pass the $3M-4$ $N$-subjettiness observables for each jet to the input layer of a fully connected DNN as the classifier. 
The resulting classifier is generally Sudakov safe~\cite{Datta:2017rhs}.
We use 3 hidden layers with between 32-512 nodes, 
each with a ReLU activation function~\cite{nair2010rectified}, followed by a sigmoid activation for the final output layer.
We train the neural network on the $N$-subjettiness observables with the Adam optimizer~\cite{Kingma2015AdamAM} and a learning rate ranging from 0.01 to 0.001 and batch size 1000, with
the binary cross entropy loss function of Ref.~\cite{https://doi.org/10.1111/j.2517-6161.1958.tb00292.x}. 
We use \texttt{Keras}~\cite{chollet2015keras}/\texttt{TensorFlow}~\cite{tensorflow2015-whitepaper} for the implementation, and determine the 
number of nodes in each hidden layer and the learning rate using a hyperparameter optimization with the Hyperband algorithm~\cite{li2017hyperband} implemented in Keras Tuner~\cite{omalley2019kerastuner}.
To crosscheck the DNN setup, we compared our results to a Random Decision Forest and found quantitatively comparable results.

\begin{figure*}[!t]
\centerline{
\includegraphics[width = 0.8 \textwidth]{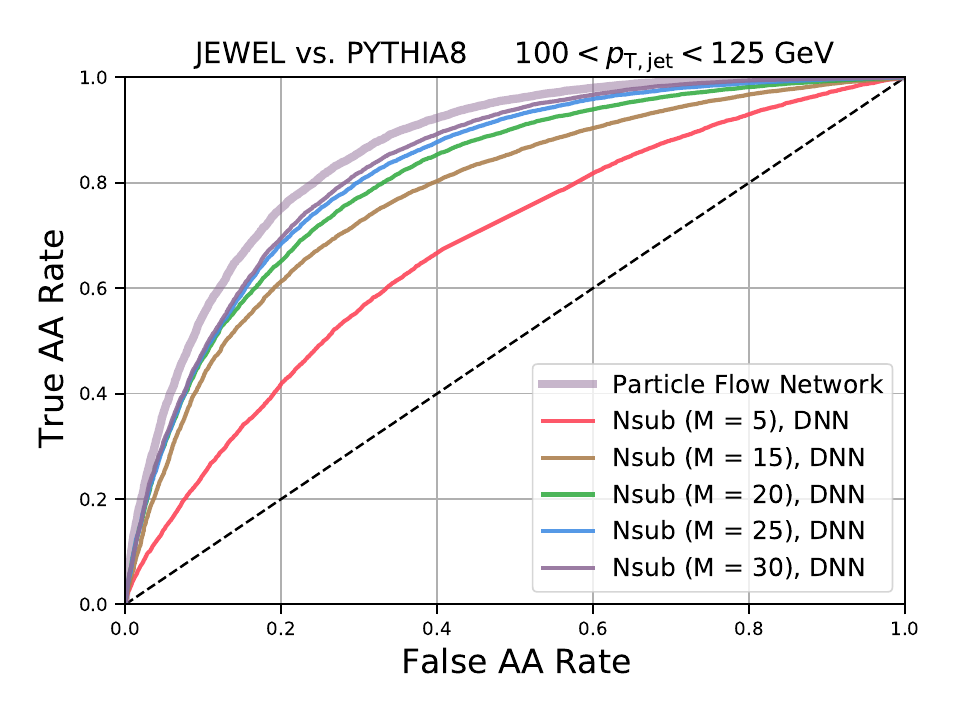}}
\caption{ROC curves for jets in $pp$ vs. $AA$ collisions using the $N$-subjettiness basis. For comparison we also show the result obtained using the classifier based on PFNs.~\label{fig:Nsubj_ROC}}
\end{figure*}

Figure~\ref{fig:Nsubj_ROC} shows the ROC curves using the $N$-subjettiness basis where we include an increasing number of $3M-4$ observables that fully specify the $M$-body phase space. 
For comparison, we also show the PFN result as in Fig.~\ref{fig:deepset_ROC} above.
While the relatively hard physics encapsulated by $M\leq5$ shows a significant discrimination power, we observe that the discrimination power significantly improves as more observables are included. Eventually, the discrimination power starts to saturate at a relatively large value of $M\approx 30$.
This indicates that soft emissions inside the jet play an important role and contain critical information about the interaction of hard probes with the QGP. 
The saturation at large $M$ found here is in stark contrast to the discrimination of QCD vs. $Z$-boson jets, as the previous studies showed saturation of the information content with the $M=4$-body phase space, i.e. a rather limited number of observables captures the complete information~\cite{Datta:2017rhs}. This has been discussed in the context of entropy production, where multiple soft emissions are produced in the jet that carry little or no additional information about the initial hard-scattering $Z$-boson signal~\cite{Neill:2018uqw}. Unlike in the case of $Z$ boson vs. QCD jets, however, for $pp$ vs. $AA$ jets we expect that there is significant information contained in the soft physics due to sensitivity to the surrounding medium in $AA$ case. 
We note that this observation is generally in agreement with the large difference between PFNs and EFNs found in Section~\ref{sec:ML_PFN}. 
Our findings suggest that it will be necessary to measure new soft-sensitive jet substructure observables in heavy-ion collisions to fully make use of the available information recorded by the experimental collaborations.
This information can be accessed by $N$-subjettiness observables for large values of $N$. We emphasize again that while the conclusions here are model-dependent, we are confident that a similar analysis can be performed with experimental data. In addition, we note that the studies here do not include the heavy-ion background, which poses a major obstacle in measurements of soft physics. We will discuss the impact of the heavy-ion underlying event in more detail in Section~\ref{sec:experiment}.

\subsubsection{Energy Flow Polynomial basis~\label{sec:EFP}}

\begin{figure}[t]
\centering
\includegraphics[width = 0.9 \textwidth]{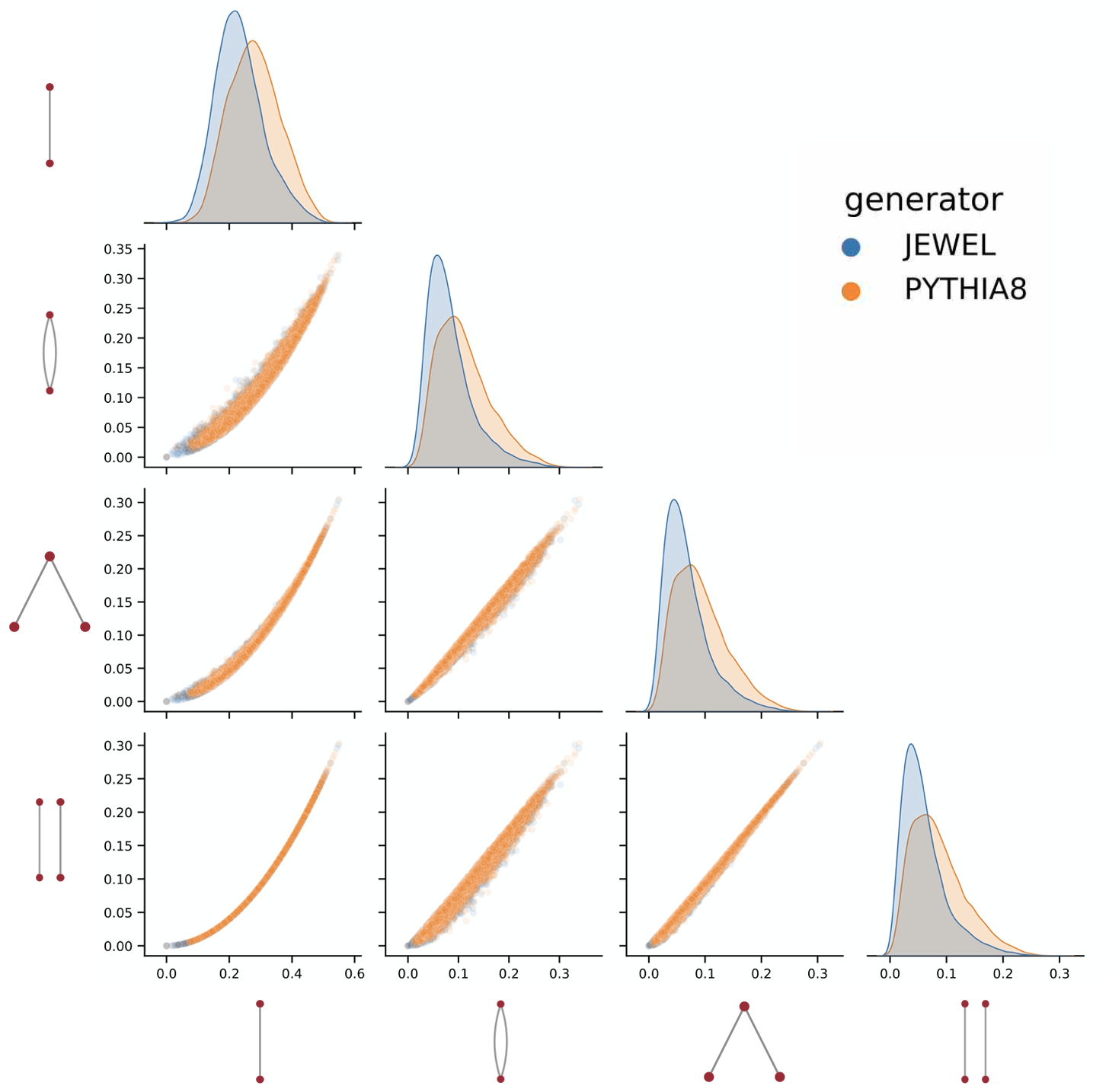}
\caption{Scatter plot showing different EFP distributions (diagonal) and their pairwise correlations (off-diagonal panels) in $pp$ and $AA$ collisions without background. The $pp$ and $AA$ results shown here are obtained from Pythia~8~\cite{Sjostrand:2007gs} and Jewel~\cite{Zapp:2012ak,Zapp:2013vla}, respectively.~\label{fig:EFP}}
\end{figure}

EFPs were introduced in Ref.~\cite{Komiske:2017aww} as an (over)complete linear basis of IRC-safe jet substructure observables. They are multi-particle correlators which can be indexed with multigraphs $G=(V,E)$ with $V$ vertices and $E$ edges. For a jet with $M$ particles, the EFP$_G$ is defined as
\begin{equation}
    \text{EFP}_G=\sum_{i_1=1}^M\cdots\sum_{i_V=1}^M z_{i_1}\cdots z_{i_V}\prod_{(k,l)\in E} \theta_{i_k i_l}\,.
\end{equation}
Here $z_i$ is a measure of the energy of particle $i$ and $\theta_{ij}$ is a pairwise distance measure of particles $i$ and $j$ in the $\eta$-$\phi$ plane
\begin{equation}
    z_i=\bigg(\frac{p_{Ti}}{\sum_j p_{Tj}}\bigg)^\kappa \,,\quad \theta_{ij}=(\Delta\eta_{ij}+\Delta\phi_{ij})^{\beta/2} \,,
\end{equation}
where $(\kappa,\beta)$ are free parameters. For $\kappa\neq 1$, the EFPs are IRC unsafe~\cite{Faucett:2020vbu}. Here, we limit ourselves to the IRC-safe EFPs, and in particular $(\kappa,\beta)=(1,0.5)$. The number of edges is the degree of the EFPs. Since the EFPs are a linear basis, any IRC-safe observable ${\cal O}$ can be approximated as
\begin{equation}
    {\cal O}\approx \sum_{G\in {\cal G}} c_G\, \text{EFP}_G\,,
\end{equation}
with coefficients $c_G$ and ${\cal G}$ is a finite set of multigraphs. 
A simple example of the EFPs is
\begin{equation}
\frac{1}{2}\;\times\;
\begin{gathered}
\includegraphics[scale=.2]{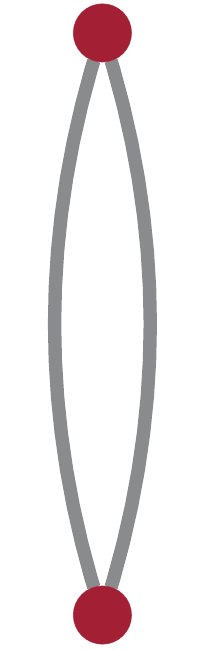}
\end{gathered} = \frac{1}{2}\sum_{i_1=1}^M\sum_{i_2=1}^M z_{i_1}z_{i_2} \theta_{i_1 i_2}^2 \,,
\end{equation}
which, for $\beta=1$, is equal to the jet mass $m_J^2/p^{{\rm jet}\,2}_T$ up to power corrections in the jet radius. The distributions and correlations of several EFPs in $pp$ and $AA$ collisions are shown in Fig.~\ref{fig:EFP}. See Refs.~\cite{Komiske:2017aww} for more details.

Since EFPs constitute a linear basis of jet substructure observables, we make use of a linear classifier.
We train a linear discriminant on the $pp$ vs. $AA$ jet data set using EFPs up to degree 7. We use an implementation with the \texttt{EnergyFlow} package~\cite{Komiske:2018cqr} and \texttt{scikit-learn}~\cite{scikit-learn}. 
Since we train a linear classifier rather than a DNN, the result is not directly comparable to the $N$-subjettiness DNN, but rather is complementary to it.

\begin{figure*}[!t]
\centerline{
\includegraphics[width = 0.8 \textwidth]{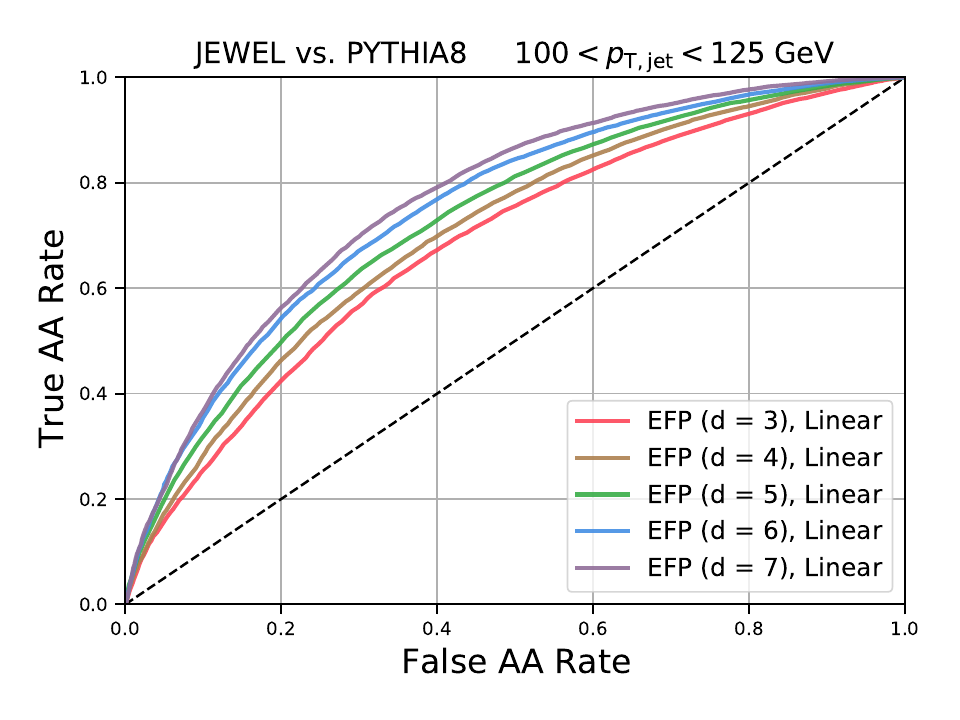}}
\caption{ROC curves for jets in $pp$ vs. $AA$ collisions using the EFP basis up to degree 7.~\label{fig:EFP_ROC}}
\end{figure*}

The corresponding ROC curves are shown in Fig.~\ref{fig:EFP_ROC}. 
We find that as we increase the EFP dimension, the classification power increases.
This is similar to the result with the $N$-subjettiness basis in the sense that a large number of EFPs needs to be included indicating that significant information is contained in the soft physics of jets.
In fact, for EFPs up to degree 7, which corresponds to 1000 observables, we do not yet observe a saturation of the discriminating power. This corroborates our results in the previous section. We note again that the performance of the different classifiers here depends on the choice of the Monte Carlo model generator. Lastly, we note that there is no unique definition of the amount of IRC-unsafe information. For example, while both the EFNs in section~\ref{sec:ML_PFN} and the linear model of EFPs here have only access to IRC-safe information, they represent different architectures and one may be trainable more efficiently than the other (leading to a better performance) depending on classification task.

\section{Observable design~\label{sec:lasso}}

The machine learning algorithms described in Section~\ref{sec:classification} deliver powerful classification ability, and themselves can be used as observables that can be measured experimentally and calculated in Monte Carlo event generators. 
However, there are limits to their ability to provide insights about jet quenching since they are generally not amenable to analytical calculations in pQCD. 
In order to address this, we propose to use machine learning to assist in \textit{observable design}, by approximating the more powerful classifiers with simpler models that are in principle calculable in perturbative QCD. 
In particular, we focus on observables with closed-form expressions that obey IRC-safety or Sudakov safety and therefore have the potential to be calculated in the future with existing methods.
To do this, we use symbolic regression to identify observables which are optimized to balance discrimination power and simplicity.
We focus on identifying single observables that maximize discriminating power, i.e.
using machine learning to identify the maximally modified jet substructure observable.
Currently, it is unknown whether the traditional jet substructure observables that have been previously measured are close to maximally modified or not;
the identification of maximally modified observables 
can provide guidance on what observables should be measured to elucidate the physics of jet quenching.
Once this machine learning observable design is done, this observable can then be measured experimentally and in principle calculated theoretically using existing methods. 
The observable design procedure can be performed either with Monte Carlo generators in order to produce an ansatz observable that can then be measured with experimental data, 
or directly with experimental data, which will be discussed in Sec.~\ref{sec:experiment}.
In the future, this machine learning approach can be extended to consider the maximum discriminating power from a set of observables, which can ultimately guide the entire program of jet substructure measurements and calculations, and motivate new input to global analyses used to extract medium properties.

\subsection{Lasso regression}

In order to select dominant features of our classifier, we use a simple form of symbolic regression known as 
Lasso (least absolute shrinkage and selection operator) regression~\cite{doi:10.1137/0907087,10.2307/2346178}.
Lasso is a linear regression model, although unlike regular linear or ridge regression, Lasso employs an $\ell_1$ regularization, which causes an increasing number of coefficients to vanish as the regularization parameter is increased.
In this way, Lasso is able to perform feature selection.
The loss function of the Lasso regression is given by
\begin{equation}
    \min_{\vec{w}} \lVert\vec y- T\vec{w}\rVert_2^2 + \lambda \lVert\vec{w}\rVert_1 \,.
\end{equation}
Here $||\cdot||_p$ denotes the $\ell_p$ norm, $\vec y$ represents the true labels, $T$ contains the input training data, $\vec{w}$ contains the weights and $\lambda$ is a regularization parameter which determines the level of sparsity of the obtained solution. For $\lambda= 0$, we recover the regular linear regression. Different than the ridge regression ($\ell_2$ regularization), the Lasso regression forces the sum of the absolute values of the coefficients in $\vec\beta$ to be less than a certain value. 

To formulate our Lasso classifier, we construct an observable that is a product of $N$-subjettiness variables, which is generally Sudakov safe. We choose the following functional form~\cite{Datta:2017lxt,Datta:2019ndh}:
\begin{equation}\label{eq:tau_K}
    {\cal O}_{N-{\rm sub}}^{\rm ML}=\Big(\tau_1^{(0.5)}\Big)^{a_{1,1}}\Big(\tau_1^{(1)}\Big)^{a_{1,2}}\Big(\tau_1^{(2)}\Big)^{a_{1,3}} \cdots \Big(\tau_{M-1}^{(1)}\Big)^{a_{M-1,1}}\Big(\tau_{M-1}^{(2)}\Big)^{a_{M-1,2}}\,.
\end{equation}
After taking the logarithm of ${\cal O}_{N-{\rm sub}}^{\rm ML}$, we can determine the $3M-4$ exponents $a_{i,j}$ as the weights of the Lasso regression.

By constructing a single observable, we provide a feasible proposal for an experimental measurement as opposed to, for example, the full set of $3M-4$ individual observables.
By measuring the product observable that is most strongly modified, one can capture the leading effect of the modification of jets in heavy-ion collisions. Moreover, choosing a Sudakov safe product form 
 is motivated by our previous findings that IRC-unsafe observables add valuable information.
This approach can be extended from a single observable to study a minimal set of observables that can achieve approximately equal discriminating power to the full set of $3M-4$ observables.
Note that despite the fact that the regression procedure requires $3M-4$ observables as input, when performing this procedure on experimental data one need not perform $3M-4$ corrected cross-section measurements. 
Rather, one can use detector-level information about these $3M-4$ observables to construct an \textit{approximate maximally discriminating observable}, which can then be measured as a single corrected cross-section. Experimental considerations will be discussed further in Sec.~\ref{sec:experiment}.

The regularization parameter $\lambda$, which can take any positive value, provides a handle to balance the performance of the classifier with the simplicity of the resulting observable.
When $\lambda$ is small, a product observable with strong classification performance but many terms will be found, and as $\lambda$ is increased, a product observable with decreased classification performance but fewer terms will be found.
That is, larger values of $\lambda$ generally decrease the number of terms, and pick out the ``most important'' $N$-subjettiness observables satisfying that regularization condition. 
When $\lambda$ becomes sufficiently large the classifier will saturate and select only one term. 
As $\lambda$ is decreased, additional terms will be added.
Note that the $N$-subjettiness terms selected by the regression procedure at one choice of $\lambda$ need not be a subset of those at a different $\lambda$. Moreover, even if they are a subset, there is nothing that requires that the ``extra'' higher-order terms to be less important -- for example the 2-term observable could be more than twice as discriminating as the 1-term observable.
The convergence of the Lasso regression can be slow for a large parameter space which is why we limit ourselves here to a relatively small number of input observables. 
For several values of $\lambda$, we find the following observables without background for $M=15$ in our Monte Carlo model studies:
\begin{align}
    \lambda=&\, 0.5: \quad {\cal O}_{N-{\rm sub}}^{\rm ML}=\tau_{14}^{(1)} \,,\\
    \lambda=&\, 0.1: \quad {\cal O}_{N-{\rm sub}}^{\rm ML}=\Big(\tau_{10}^{(1)}\Big)^{0.071} \Big(\tau_{11}^{(1)}\Big)^{0.157} \Big(\tau_{14}^{(1)}\Big)^{0.649} \tau_{14}^{(2)}\,,\\
    \lambda=&\, 0.01: \quad {\cal O}_{N-{\rm sub}}^{\rm ML}= \Big(\tau_{2}^{(0.5)}\Big)^{0.608} \Big(\tau_{4}^{(2)}\Big)^{-0.186} \times...\times \tau_{14}^{(2)}\; \quad\mathrm{(23\;terms)} \,.
\end{align}
Since we can rescale the exponents by an overall factor without changing the performance of the classifier, we choose the exponent of the rightmost factor in Eq.~\ref{eq:tau_K}, in this case $\tau_{14}^{(\beta)}$, as 1 for all values of $\lambda$.

\begin{figure*}[!t]
\centerline{
\includegraphics[width = 0.8 \textwidth]{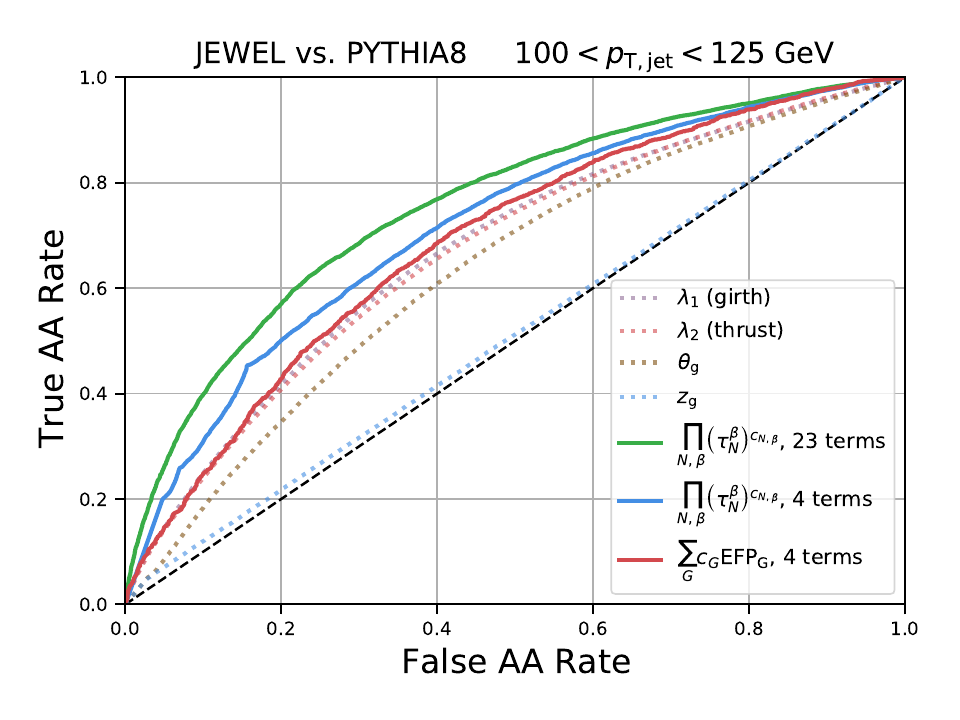}}
\caption{ROC curves for the Lasso regression using the $N$-subjettiness basis and EFPs. For comparison we also show the result for typical observables in heavy-ion collisions.~\label{fig:Lasso}}
\end{figure*}

We find that the Lasso regression generally prefers large values of $N$. 
For sufficiently large values of $\lambda$, we find that the Lasso regression always picks only one observable which turns out to be one of the $N$-subjettiness observables with the largest allowed value of $N$.
When $\lambda$ is lowered gradually, the Lasso regression adds additional $N$-subjettiness observables with intermediate values of $N$. If we further lower $\lambda$, the Lasso regression eventually adds observables with small values of $N$. This is generally in line with our previous observations that significant information about jet quenching is in the soft physics. 
While large-$N$ $N$-subjettiness are not simple to calculate, our results suggest
motivation to extend perturbative QCD calculations to increasingly large $N$.
Interpreting the exact jet quenching physics message from these product observables will require more theoretical work – here we focus mainly on the important step that these are in principle calculable, unlike the high-dimensional machine learning inputs.
We emphasize that these results are model-dependent, and the procedure should be repeated on experimental data in order to obtain nature's maximally discriminating single observables.
We also compared our results to regression algorithms with an $\ell_0$ regularization (NP-hard) and found similar results.
The exact set of observables which is chosen can vary somewhat when we run the Lasso regression with a different random seed. This is likely due to the existence of multiple local minima and the shared information between different $N$-subjettiness observables, see also Fig.~\ref{fig:nsubjettiness}. We also find that the discrimination power increases only gradually as we allow for more observables to be included in the product observable. This observation is in line with the findings of Ref.~\cite{Larkoski:2014pca}. We expect that more general symbolic regression techniques as developed for example in Ref.~\cite{cranmer2020discovering} can also be adapted to identify suitable jet substructure observables which we plan to explore in the future work. See also Ref.~\cite{Butter:2021rvz}.

\begin{figure*}[!t]
\includegraphics[width = 0.5 \textwidth]{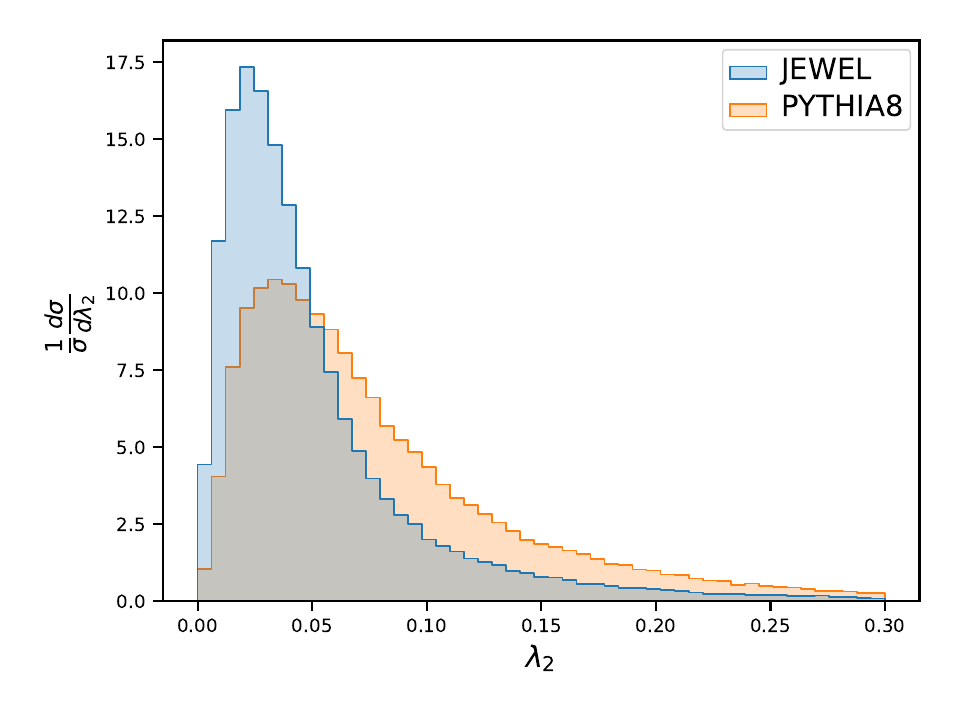}
\includegraphics[width = 0.5 \textwidth]{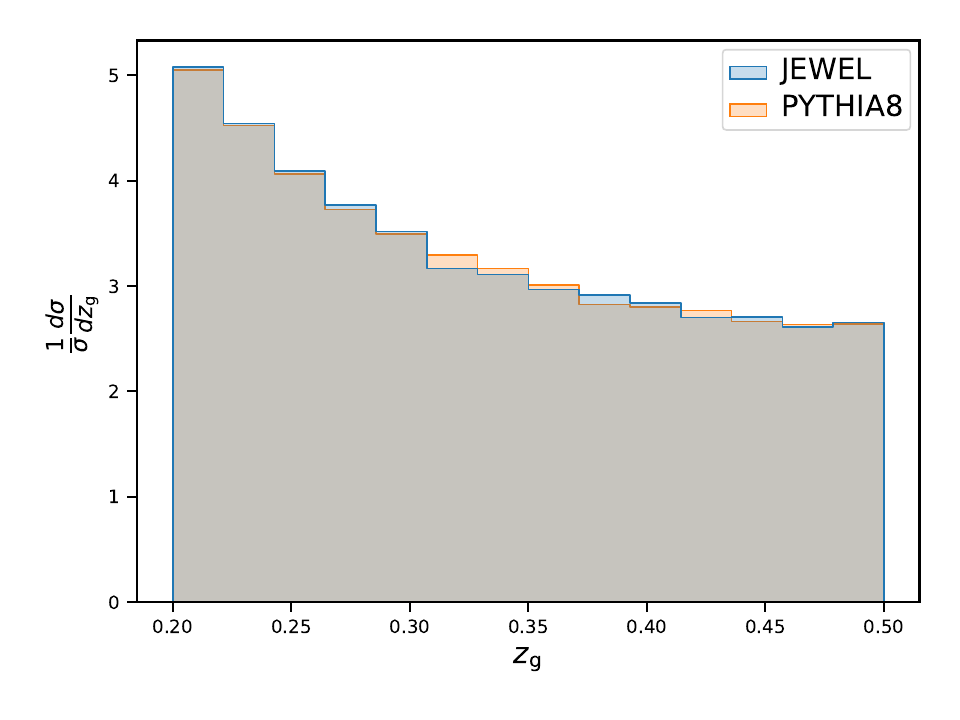}
\includegraphics[width = 0.5 \textwidth]{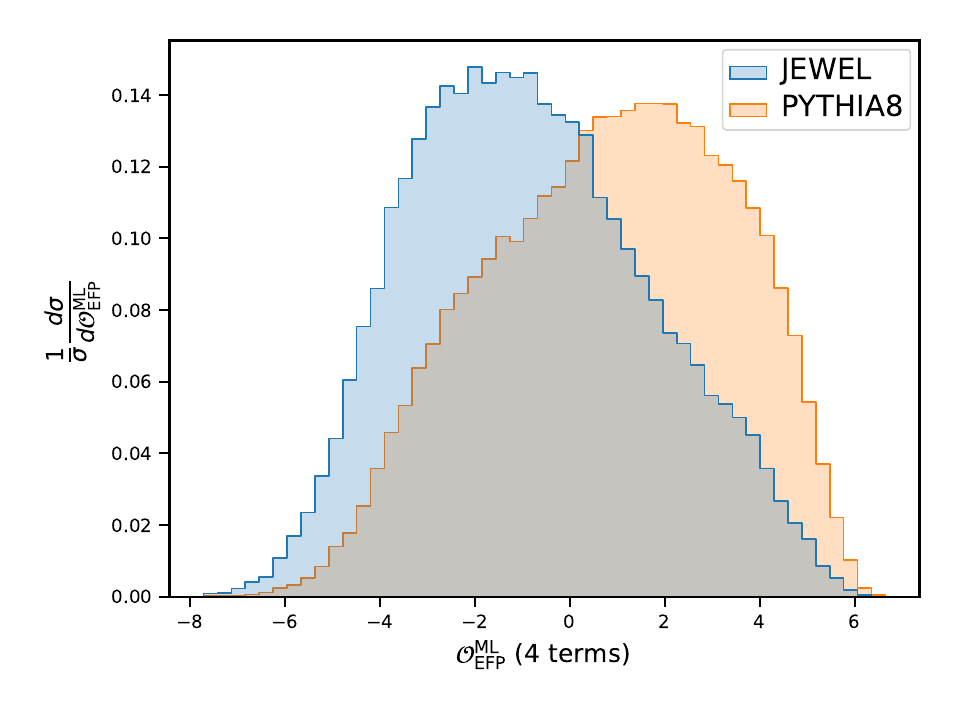}
\includegraphics[width = 0.5 \textwidth]{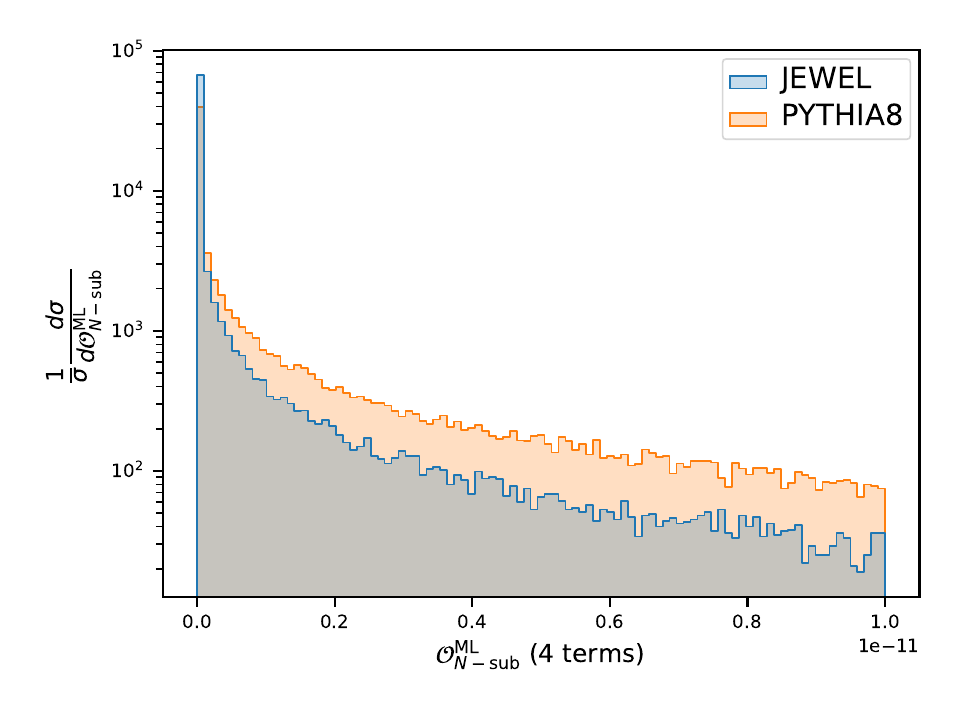}
\caption{Distributions of observables in $pp$ and $AA$ collisions which have already been measured by experimental collaborations and examples of the machine-learned observables using the $N$-subjettiness and EFP basis.~\label{fig:observable_distributions}}
\end{figure*}

To illustrate the performance of the different observables, we show the corresponding ROC curves in Fig.~\ref{fig:Lasso}. As expected, solutions with more terms (lower values of $\lambda$) lead to better discriminating power at the expense of an increased complexity of the observable.
For comparison, we also plot the ROC curves of four observables which have already been measured in heavy-ion collisions: the soft drop momentum sharing fraction $z_g$~\cite{Larkoski:2015lea,Cal:2021fla, CMS:2017qlm, ALICE:2021obz}, the groomed jet radius~\cite{Larkoski:2014wba,Kang:2019prh, ATLAS:2019mgf, ALICE:2021obz, STAR:2021kjt}, jet thrust~\cite{ALICE:2017nij, CMS:2018fof} and jet girth~\cite{ALICE:2018dxf}. We observe that the machine-learned product observables outperform these observables. The distributions of these observables are shown in Fig.~\ref{fig:observable_distributions}.
Note that the $N$-subjettiness product observable distribution exhibits a large difference between JEWEL and PYTHIA8 for ${\cal O}_{N-{\rm sub}}^{\rm ML}=0$. This indicates that a significant amount of the classification power is related to the multiplicity of the jet, since $\tau_N^{\beta}$ vanishes when the jet constituent multiplicity is $<N$. This aspect of the modification is highly model-dependent, however, and the exact ${\cal O}_{N-{\rm sub}}^{\rm ML}$ found here should be taken for illustrative purposes only, and should be determined using experimental data.

In addition to an $N$-subjettiness product observable ansatz, we consider a linear combination of EFPs.
As discussed above, EFPs form a linear basis of jet substructure observables making them ideally suited for a Lasso regression~\cite{Komiske:2017aww}. Here we preprocess the input data to zero mean and unit variance.
Analogous to the previous section, we determine the coefficients $c_G$ of
\begin{equation}
    {\cal O}_{\rm EFP}^{\rm ML} = \sum_{G\in {\cal G}} c_G\, \text{EFP}_G\,,
\end{equation}
except that now we use a Lasso regression instead of a ``regular'' linear regression which will pick out only the most important terms. As an example, we use EFPs up to degree 4 and the Lasso regression with $\lambda=0.001$, for which we find a very stable result:
\begin{equation}\label{eq:lasso_efp}
\begin{gathered}
\includegraphics[scale=.2]{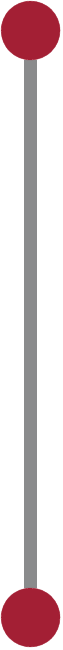}
\end{gathered}
\; + \; 3.54\;\times\;
\begin{gathered}
\includegraphics[scale=.2]{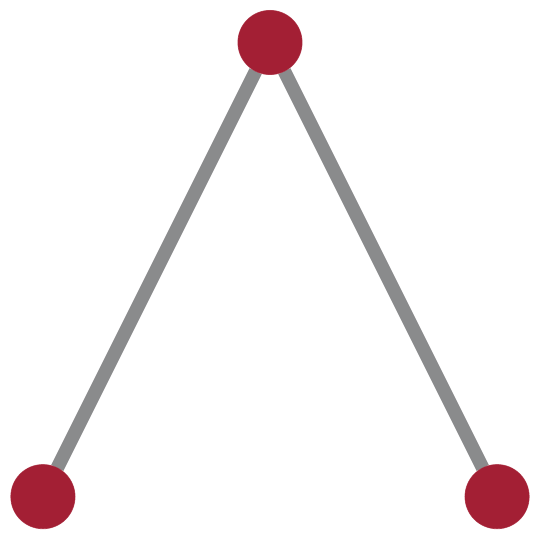}
\end{gathered}
\; +\; 1.72\;\times\;
\begin{gathered}
\includegraphics[scale=.27]{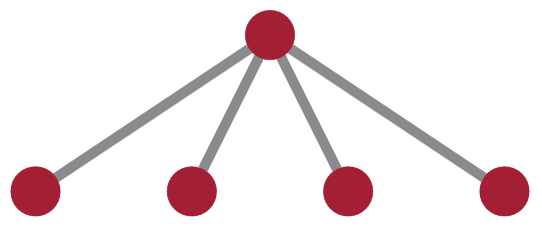}
\end{gathered}
\; -\; 3.82\;\times\;
\begin{gathered}
\includegraphics[scale=.2]{2_2_1.pdf}
\includegraphics[scale=.2]{2_2_1.pdf}
\end{gathered}\,.
\end{equation}
The corresponding ROC curve and the distribution of this ML-learned observable are shown in Figs.~\ref{fig:Lasso},~\ref{fig:observable_distributions}, respectively. We find that despite the simplicity of the machine-learned EFP observable, it outperforms the other ``traditional'' observables. The intriguing aspect of observables which involve a relatively small number of EFPs, as in Eq.~(\ref{eq:lasso_efp}), are that they are generally calculable within perturbative QCD, see Ref.~\cite{Cal:2022fnm} for recent progress. Moreover, we note that some combinations of EFPs are easier to calculate directly compared to individual EFPs~\cite{Komiske:2017aww}. The same holds for certain products of $N$-subjettiness observables~\cite{Datta:2017rhs}. We leave the identification of the subspace of observables that is relatively simple to calculate within perturbative QCD and the combination with ML techniques like the Lasso regression for future work.

\section{Information loss: the underlying event and background subtraction
~\label{sec:background}}

The large, fluctuating underlying event produced by the 
QGP causes notorious experimental and theoretical challenges in heavy-ion collisions -- in particular, by limiting which observables can be reliably measured.
Typically, background subtraction procedures are applied in order to mitigate this problem. Systematic uncertainties associated with the subtraction are estimated in order to adequately capture the lack of exact knowledge of which particles arise from the underlying event, and which from the jet.

From the perspective of information content, this presents two distinct mechanisms
by which the information in jet quenching can be lost.
First, the fluctuating underlying event can be viewed as a source of noise. One cannot distinguish particles arising from underlying event from those correlated to the jet, and so to the extent that the noise distribution overlaps with the signal distribution, the ability to distinguish the two is irrecoverably reduced.
Second, background subtraction algorithms themselves can cause information loss. Since background subtraction inherently involves removal of particles from the jet, and one does not have exact knowledge of which particles arise from the underlying event, this procedure strictly results 
in information loss.

\begin{figure*}[!t]
\centerline{
\includegraphics[width = 0.8 \textwidth]{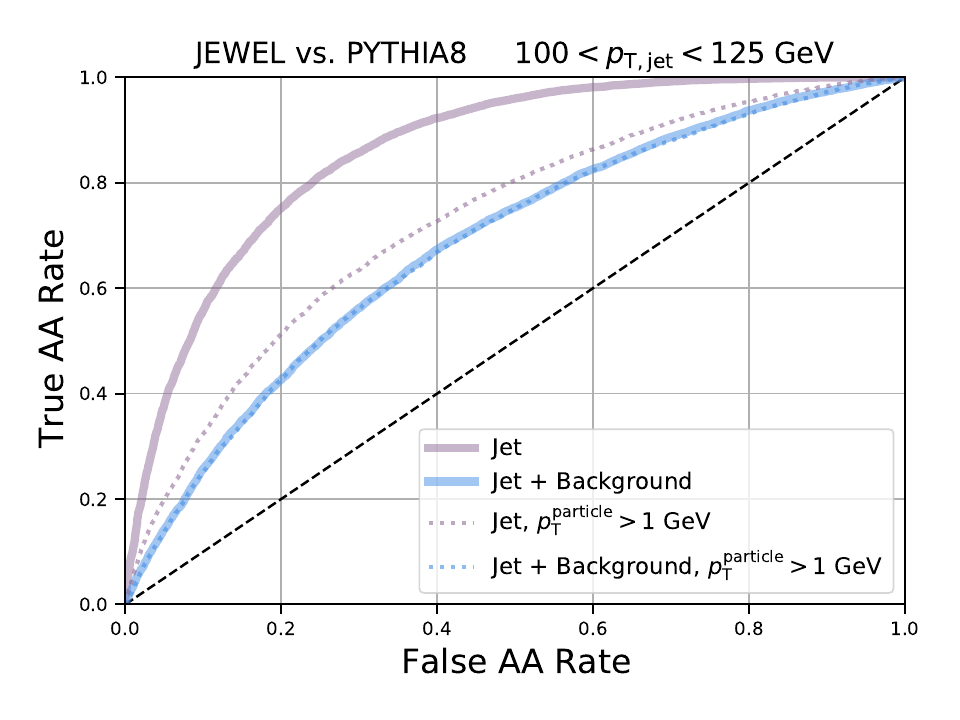}}
\caption{ROC curves for PFNs trained with (i) PYTHIA8/JEWEL jets, 
(ii) jets clustered from a combination of PYTHIA8/JEWEL events with a thermal background, with event-wide constituent subtraction applied ($\Rmax=0.25$), (iii) PYTHIA8/JEWEL jets only considering jet constituents with $\pT>1$~GeV,
and (iv) jets clustered from a combination of PYTHIA8/JEWEL events with a thermal background, only considering jet constituents with $\pT>1$~GeV, with event-wide constituent subtraction applied ($\Rmax=0.25$). ~\label{fig:roc-background}}
\end{figure*}

The jet classification methods used in Section~\ref{sec:classification} can be used to evaluate the magnitude of each of these contributions. 
Within the context of the parton shower models considered, we assess the overall impact of the underlying event on the jet classification performance by comparing a PFN trained only on the hard jet particles to a PFN trained on the combination of jet and background particles (after performing constituent subtraction, as described in Sec.~\ref{sec:event_generation}).
In order to perform a controlled comparison, 
in both cases we select jets based on the \pTjet{} of 
the particles from PYTHIA8 or JEWEL alone.
Figure \ref{fig:roc-background} shows that there is a dramatic decrease in the classification power due to the presence of the underlying event. 
We also plot PFNs trained on jet particles with $\pT>1$~GeV.
Comparing the ROC curves with and without this requirement, we find that in the case without background, a large discrimination power resides in the soft physics – whereas in the case with background, the presence of soft information makes no difference.
That is, in the presence of background, sufficiently soft discrimination is no longer useful – and the discrimination is dominated by hard physics. 
This observation presents a delicate challenge for the study of jet quenching: soft information is crucial to maximize discrimination between quenched and unquenched jets,
yet the fluctuating underlying event fundamentally prevents much of this information from being accessed.

In order to disentangle how much of this information loss is due to inherent noise of the fluctuating background, and how much is due to the background subtraction algorithm, we compare the classification performance before and after background subtraction is performed. To do this, we fix the jet axis using the hard jet in order to avoid distortion of the axis by the fluctuating background. We consider three different jet populations, each defined by a cone of particles within rapidity-azimuth distance of $\Delta R=0.4$ from the common axis. The three different particle selections we consider are:
\begin{enumerate}
    \item Hard jet particles only
    \item Hard jet particles and background particles, with constituent subtraction applied
    \item Hard jet particles and background particles, without background subtraction
\end{enumerate}

\begin{figure*}[!t]
\centerline{
\includegraphics[width = 0.7 \textwidth]{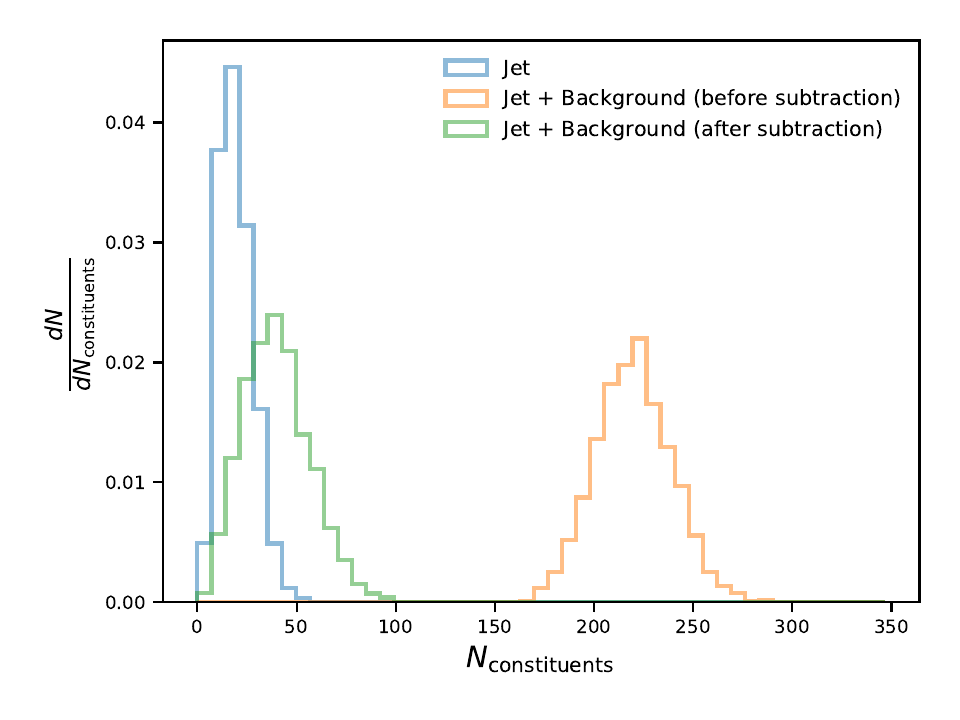}}
\caption{Distributions of particle multiplicities in the training jet population.~\label{fig:cs-multiplicity}}
\end{figure*}

To illustrate the differences between these three selections, note that the background subtraction procedure can cause a dramatic change in the particle multiplicity of the jet. Figure \ref{fig:cs-multiplicity} shows the
distribution of multiplicities of jet constituents before and after background subtraction, for the nominal constituent subtraction configuration with $\Rmax=0.25$. Application of the background subtraction procedure brings the constituent multiplicity distribution much closer to the distribution in the absence of background. The dramatic change, however, emphasizes the possibility that the background subtraction procedure may cause significant information to be lost.
Given the increased number of particles before background subtraction, when training the PFN before background subtraction, we double the number and size of all layers, as well as the dimension of the latent space, which we observe to give an improvement to the PFN performance (the case after background subtraction and the hard jet are unaffected).

\begin{figure*}[!t]
\centerline{
\includegraphics[width = 0.8 \textwidth]{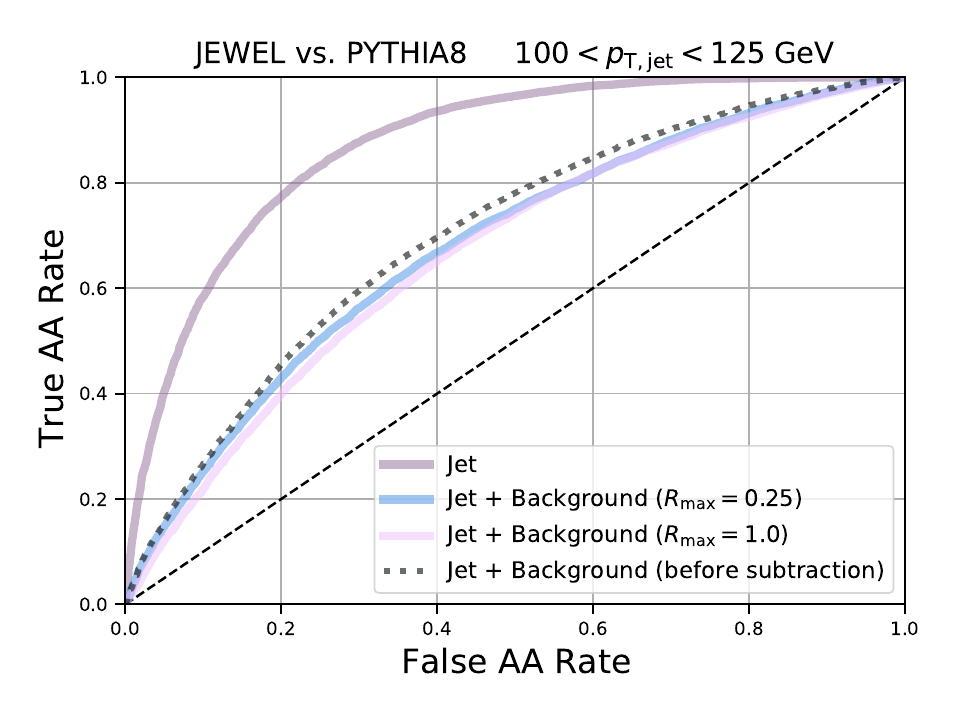}}
\caption{ROC curves comparing the performance with (i) Hard jet particles only, (ii) Hard jet particles and background particles, with constituent subtraction applied (for two different values of \Rmax{}, and (iii) Hard jet particles and background particles, without any background subtraction applied~\label{fig:roc-cs}}
\end{figure*}

Figure \ref{fig:roc-cs} shows the ROC curve for PFNs trained before and after background subtraction (for two values of \Rmax{}), with comparison to the hard jet for reference.
The ROC curve before background subtraction is higher than
that after background subtraction, indicating that, within our model studies, the constituent subtraction procedure does remove a small but significant amount of information. As the background subtraction becomes stronger, going from $\Rmax=0.25$ to $\Rmax=1.0$, we observe additional loss of information.
This constitutes the first study quantifying the information loss of background subtraction algorithms.
We propose that the ROC curve should be used as a metric to quantify the information loss, and can be used to tune optimal parameters in the algorithm – including machine learning based background subtraction~\cite{Haake:2018hqn}.

These results come with the caveat that there is a third source that can deteriorate the performance of the classifier in the presence of background: the performance of the PFN may eventually deteriorate due to the increased size of the training data.
In particular, we note that it is possible for the ROC curve to improve after the background subtraction but only if the PFN is not a sufficiently good approximation~\cite{DBLP:journals/corr/abs-1901-09006} of the optimal classifier which is given by the likelihood ratio according to the Neyman-Pearson lemma~\cite{Neyman:1933wgr}.
Nevertheless, we can unambiguously see that, within our model studies, the background subtraction procedure (which \textit{reduces} the size of the per-jet training data) overall deteriorates the performance of the PFN.
 
\section{Application to experimental data~\label{sec:experiment}}

In order to assess the actual information content that distinguishes jets in $AA$ collisions from those in $pp$, our analysis can be performed using the event-by-event data collected by the experimental collaborations at the LHC and RHIC. In this section we outline several considerations for how such an analysis can be carried out. 

In experimental data, one cannot unambiguously separate the hard jet component from the QGP underlying event. Consequently, the quenched jets in heavy-ion collisions should be compared to a sample of proton-proton jets embedded into heavy-ion background, with or without background subtraction. These two samples define the two classes to be distinguished by the classifier, which we denote $AA$ and $pp\,\oplus AA$, respectively.

The simplest approach to measure the ROC curve experimentally is to train the classifier on detector-level quantities: $AA$ vs. $pp\,\oplus AA$, taking careful consideration 
of differences in detector effects in the two samples.
Note that experimentally, one does not have access to the ``hard jet'' \pTjet, unlike the studies we performed in Sec.~\ref{sec:background}. Instead, 
one can only select jet populations based on the \pTjet{} in the presence of background, for example after background subtraction.
In order to correct for the detector and the background fluctuation effects, one can apply corrections either
after training (such as with a bin-by-bin correction) or before training (with an unfolding procedure to correct the inputs).
Using traditional unfolding methods, however, one cannot easily build a response matrix to
perform the correction, unless one performs an unfolding of the high-dimensional input space, such as a $3M-4$-dimensional set of $N$-subjettiness observables.
With the recent prospect of machine learning based methods to unfold high dimensional events ~\cite{Andreassen:2019cjw}, this may be possible, even to the extent of unfolding full events and thereby enabling the training of a classifier directly on corrected particles.

There are several additional challenges in performing these measurements compared to the Monte Carlo studies presented above. First, the detector conditions between the proton-proton and heavy-ion data taking periods may be different – and the classifier will naively learn these differences. Second, in the $pp\,\oplus AA$ jet sample, one must ensure that only soft particles – and not hard jets – enter the distribution from the embedded heavy-ion event.
Third, the size of the jet sample is limited by the available statistics recorded by the experiment, which in turn can limit the performance of the classifier.
These challenges are each surmountable, and we are optimistic that such an analysis can be performed at the LHC.

We propose that each of the three complementary studies in Sections~\ref{sec:classification}-\ref{sec:background} can be performed on experimental data:
\begin{itemize}
    \item \textit{Measuring the ROC curve}. The measured ROC curve can serve as an observable that can be compared to Monte Carlo event generators. Moreover, the distribution of information content with complete sets of jet substructure observables can provide a differential test of jet quenching models, to the extent that highly soft-sensitive observables, such as high-$N$ $N$-subjettiness or high-dimension EFPs, can be reliably measured in the presence of the heavy-ion underlying event.
    \item \textit{ML-assisted observable design}. Regardless of whether the classifier is trained on detector-level inputs or corrected inputs, symbolic regression can be used to identify approximate maximally discriminating observables. These identified observables can then be measured with traditional techniques: correcting for detector and background effects, and in principle comparing to jet quenching calculations. 
    \item \textit{Information content and background subtraction techniques}. The information loss caused by various background subtraction algorithms can be quantified by comparing classification performance before and after subtraction, and can be used to select and tune subtraction algorithms to minimize information loss.
\end{itemize}

\section{Conclusions and outlook~\label{sec:conclusions}}

We explored the use of machine learning classifiers to investigate the information content distinguishing jets in heavy-ion collisions from jets in proton-proton collisions.
The labels for the fully supervised machine learning task here are unambiguous and we expect that new insights about the nature of the quark-gluon plasma (QGP) can be obtained with our approach. As a proof of concept, we trained classifiers and designed new observables using jet samples generated from parton shower event generators. We expect that the developed techniques can be applied directly to event-by-event experimental data.

Within our model studies, 
we found that there is a significant ability to distinguish jets in the absence of the heavy-ion underlying event.
By comparing Infrared Collinear (IRC) unsafe Particle Flow Networks to IRC safe Energy Flow Networks, we found that a substantial amount of this information resides in IRC-unsafe physics. 
By using complete sets of IRC safe $N$-subjettiness observables and Energy Flow Polynomials, we studied the convergence of the classifier as a function of the number of observables included in the training procedure, and found that the performance saturates only when a large number of observables are included.
This demonstrates that a substantial amount of information is contained in soft emissions inside the jet, and is in stark contrast to e.g. QCD vs. $Z$-jet tagging where only a small number of observables is sufficient.

The ability to distinguish jets in heavy-ion collisions from jets in proton-proton collisions appears to substantially decrease in the presence of the heavy-ion underlying event. 
Our model studies indicate that this is primarily due to fluctuations inherent to the background, which lead to irrecoverable information loss, but that there is also a significant information loss due to commonly used background subtraction algorithms. We performed the first such study assessing the information loss of background subtraction, which can be used to tune and develop new background subtraction algorithms in the future to preserve the information content of the jet quenching process.

In order to directly connect our machine learning approach to theoretical calculations, we designed new observables using Lasso regression that maximize the discrimination power between jets in heavy-ion collisions and jets in proton-proton collisions and are in principle calculable in perturbative QCD. 
By finding observables that maximize a balance of being (i) highly discriminating, and (ii) in principle calculable, we provide guidance for the theoretical understanding of jet quenching.
These studies can be extended in the future to not just explore maximally modified observables, but to identify sets of observables that offer the maximal ability to constrain QGP properties. Already, our observation that the soft physics of jets plays an important role provides motivation to study specifically soft-sensitive jet substructure observables – which may bridge the gap between traditional jet substructure measurements and the machine-learned classifiers.
We expect these methods to inform future experimental analyses and theoretical focus.

It is essential that the studies presented here be performed on event-by-event experimental data,
since parton shower event generators for jet quenching involve assumptions ranging from the role of medium response to the factorization structure of jet quenching.
Since the training labels are exactly known, this serves as a case in which machine learning can be used directly on experimental data without reliance on modeling.
The studies should be extended to different jet topologies, different kinematic ranges, and multiple jet radii.
Such studies will present new opportunities to use jets to precisely constrain QGP properties,
and can reveal the extent to which
the highly discriminating soft information encoding jet quenching is irrecoverably lost due to the heavy-ion underlying event.
Lastly, we note that while our studies here focus specifically on jet substructure, we expect that similar techniques can be applied to full events in heavy-ion collisions or in electron-nucleus collisions at the future Electron-Ion Collider.

\section*{Data availability}

The data sets used in this work can be found at: \\

\href{https://zenodo.org/record/5758081\#.Ya_PAS2cZTY}{\tt 10.5281/zenodo.5758081} \\

\noindent
We provide the particle four vectors and $N$-subjettiness values of jets in $pp$ (Pythia~8~\cite{Sjostrand:2007gs}) and $AA$ (Jewel~\cite{Zapp:2012ak,Zapp:2013vla}) collisions for the kinematics specified above. The data sets with and without the thermal background are included.

\section*{Acknowledgments}

We would like to thank Raghav Elayavalli, Andrew Larkoski and Ben Nachman for helpful discussions and Ben Nachman for feedback on our manuscript. YL, JM, MP are supported by the U.S. Department of Energy, Office of Science, 
Office of Nuclear Physics, under the contract DE-AC02-05CH11231. YSL, MP and FR
were supported by the LDRD Program of Lawrence Berkeley National Laboratory. FR is supported by the Simons Foundation under the Simons Bridge program for Postdoctoral Fellowships at SCGP and YITP, award number 815892 and the NSF, award number 1915093.

\medskip


\bibliographystyle{JHEP}
\bibliography{main}

\end{document}